\documentclass[preprint,APS,showpacs,preprintnumbers,amsmath]{revtex4}
\usepackage{graphicx}
\usepackage{dcolumn}
\usepackage{bm}
\usepackage{wrapfig}

\def\be{\begin{equation}}
\def\ee{\end{equation}}
\def\e#1{\label{#1}\end{equation}}
\def\bea{\begin{eqnarray}}
\def\eea{\end{eqnarray}}
\def\ea#1{\label{#1}\end{eqnarray}}

\def\bem#1{\begin{mathletters}\label{#1}}
\def\eml{\end{mathletters}}

\def\4#1{{\boldsymbol{#1}}}
\def\8#1{{\widetilde{#1}}}
\def\bse{\begin{subequations}}
\def\ese{\end{subequations}}
\def\nn{\nonumber}
\begin{document}
\title{From Zeno to anti-Zeno: decoherence-control dependence on the quantum statistics of the bath}
\author{D. D. Bhaktavatsala Rao}
\email{dasari@weizmann.ac.il}
\author{Gershon Kurizki}
\affiliation{Weizmann Institute of Science, Rehovot 76100, Israel.}%
\begin{abstract}
We demonstrate through exact solutions that a spin bath leads to stronger (faster) dephasing of a qubit than a bosonic bath with identical bath-coupling spectrum. This difference is due to the spin-bath ``dressing'' by the coupling. Consequently, the quantum statistics of the bath strongly affects the pulse sequences required to dynamically decouple the qubit from its bath. 
\end{abstract}
\date{\today}
\maketitle

\section{Introduction}
An essential requirement for realizing fault-tolerant quantum computation is the ability to protect the information encoded in the qubits from leaking into the environment (bath) \cite{nielsen}. The stronger the qubit-bath coupling, the faster is the loss of information (decoherence) \cite{breu}. The desirable but often unachievable goal is to completely decouple the system from the bath. Yet, even if they cannot be completely eliminated the adverse effects of decoherence suppressed using control pulses at a rate faster than the inverse memory (correlation) time of the bath \cite{cc,koff}.  For a common kind of decoherence known as pure dephasing \cite{cc} the problem can be solved exactly in certain cases \cite{uhrig1}. If the bath is taken to be bosonic, i.e., composed of harmonic oscillators, but otherwise general, its control can be analyzed by a second-order non-Markovian master equation \cite{koff}. This master equation has yielded a formula for optimal dephasing control \cite{goren} which has been extended to arbitrary forms of decoherence \cite{jens}. Interestingly, similar results are reproduced when the source of dephasing is assumed to be classical noise, whose power spectrum coincides with the coupling spectrum of the quantum bath \cite{uhrig1,goren}. 

There has been a growing interest in decoherence control from fermionic (spin) baths \cite{loss,gaa,das,dal} as they form the main source of decoherence for solid-state qubits \cite{spbath}. Decoherence in these systems has generally been treated by classical models: for example, in the quasi-static approximation \cite{merk,petta}, only the (effective) magnetic field distributions caused by the bath spins determine the dephasing rates and the optimal spin-echo times.

In general, dynamical control methods set out to affect the qubit-bath dynamics on a time-scale much shorter than the bath correlation (memory) times known as the Zeno regime \cite{azenat}. On this time scale, the qubit-bath interaction dynamics is coherent and reversible \cite{azenat}. The control then moves the qubit's frequency away from the harmful bath modes, thus approaching dynamical decoupling \cite{cc}. When the control is instead applied in the anti-Zeno regime, i.e.,on the time scale comparable to the correlation time, it can drastically enhance the decoherence rate \cite{pascas, azenat}. Hence, the knowledge of the cross-over from the Zeno to the anti-Zeno regime is crucial for designing pulse sequences to control decoherence. 

The foregoing studies leave us with some important questions: Is the knowledge of the bath (or noise) spectrum sufficient to design optimal control ? Does one need a rigorous quantum analysis of the system-bath dynamics  or can it be substituted by classical-noise analysis \cite{uhrig1,goren} ? How does the bosonic or fermionic nature of the bath constituents influence the time scales of dephasing and its control ? Recent works have addressed the differences between the spin and bosonic baths using second order master equations \cite{comp1,comp2}. Here we exactly analyze pure dephasing and its control, exactly for spin and bosonic baths and contrast, the two, showing the limitations of second-order master equations for each bath.

We point out that pure dephasing of a qubit coupled to a spin bath, cannot in general be treated within the classical noise description (e.g., by assuming an effective magnetic-field distribution \cite{merk,petta}) as opposed to dephasing caused by a bosonic bath. We demonstrate that even if the coupling spectra of the two kinds of baths are taken to be identical, the resulting dephasing is drastically different. In particular, we show that the Zeno to anti-Zeno crossover is not solely determined by the bath coupling spectrum \cite{koff,azenat}, but strongly depends on the quantum statistics of the bath constituents.  Correspondingly, the same pulse rate is shown to fall within the Zeno (decoherence-suppressing) regime for one bath and within the anti-Zeno (decoherence-enhancing) regime for the other. 

\section{Qubit dephasing}
The coupling of the qubit (S) to the spin-bath ($\sigma$) is described by the qubit-spin (spin-spin) Hamiltonian
\be
\label{hams}
H_{S\sigma} = \Delta(t) S^x + \sum_k \omega_k \sigma^k_z + \sum_k \eta_k S^z (\sigma^+_k+\sigma^-_k).
\ee
The first term represents the control Hamiltonian for the qubit $(\vec{S})$ with strength $\Delta(t)$, $\omega_k$ represents the (Zeeman) energy of the $k^{th}$ bath spin ($\vec{\sigma}_k$), and $\eta_k$ is the corresponding coupling strength of the spin to the qubit. The control and the bath act on perpendicular Bloch-sphere axes of the qubit $x$ and $z$ respectively.

Similarly, when the qubit is coupled to a bosonic bath with identical bath coupling spectrum, its evolution is governed by the spin-boson Hamiltonian \cite{breu}
\be
\label{hamb}
H_{Sb} = \Delta(t) S^x + \sum_k \omega_k b^\dagger_kb_k + \sum_k \eta_k S^z (b^\dagger_k+b_k),
\ee
where $b$, $b^\dagger$ represent the creation and annihilation operators of the $k^{th}$ boson (oscillator).

\subsection{Pure dephasing without control}
\noindent
(i) {\it Spin bath}:
In the absence of control, the system-bath evolution leads to pure dephasing of the qubit. In contrast to dephasing by a bosonic bath (App. A) the Magnus expansion in the case of a spin bath does not truncate to any order (App. A). Hence, obtaining the time evolution operator by this procedure is quite cumbersome.

In the $S^z$ basis of the system ($|\mathcal{+}\rangle, |{\bf -}\rangle$), the Hamiltonian of a qubit in a spin bath Eq. (\ref{hams}) with $\Delta(t)=0$, can be rewritten as 
\bea
H_{S\sigma} &=& H^+_\sigma|\mathcal{+}\rangle\langle +| + H^-_\sigma|{\bf -}\rangle\langle-|, \nn \\
H^{\pm}_\sigma &=& \sum_k \omega_k \sigma^k_z \pm \sum_k \eta_k (\sigma^+_k+\sigma^-_k)
\eea 
where $H^{\pm}_\sigma$ are the bath operators. 
The simple form of $H^{\pm}_\sigma$ makes it easy to diagonalize, so that we obtain a closed-form equation for the time-evolution operator
\be
\label{uop}
U_{\sigma}(t) =\left[U^+_\sigma(t)|\mathcal{+}\rangle\langle +| + U^-_\sigma(t)|{\bf -}\rangle\langle-|\right]
\ee
where
\be
\label{udef}
U^\pm_{\sigma}(t) = \prod_k U^\pm_{\sigma,k}(t), ~~ U^\pm_{\sigma,k}(t) = \left [\cos \delta_k t \hat{\rm I}+ \frac{i\sin \delta_k t}{\delta_k}(\omega_k \sigma^z_k \pm \eta_k \sigma^x_k)\right].
\ee
Here the renormalized eigenfrequencies of the bath spins are $\delta_k = \sqrt{\omega_k^2 + \eta_k^2}$. As shown below, this renormalization would lead to faster dynamics for a spin-bath than a bosonic-bath. 

As the above dynamics leads only to pure dephasing of the qubit, the populations of the $|+\rangle$ and $|-\rangle$ states are unchanged, while the off-diagonal elements $\rho^{+-}_S(t)$
dynamically evolve in time as
\be
\label{coh}
\rho^{+-}_S(t)={\rm e}^{-t\Gamma_\sigma(t)}\rho^{+-}_S(0), ~~ \Gamma_\sigma(t) = -\frac{1}{t}\ln\left(Tr_B[U^+_B(t)\rho_\sigma(0)U^-_B(t)]\right).
\ee
where $\rho_\sigma(0)$ is the initial state of the bath spins.

We consider an uncorrelated initial state of the bath spins $\rho_\sigma = \frac{1}{\mathcal{Z}}\exp(-\beta H_B)$. In the energy basis, the state of a $k^{th}$ bath-spin
can be written as $\rho_{\sigma,k} = p_k|e\rangle\langle e| + (1-p_k)|g\rangle\langle g|$, where $p_k = \exp(-\beta\omega_k)/2\cosh\beta\omega_k$. 
Since the trace of bath operators obeys the simple symmetry rule $Tr[U^+_{\sigma,k}(t)|e\rangle\langle e|U^-_{\sigma,k}(t)]$ = $Tr[U^+_{\sigma,k}(t)|g\rangle\langle g|U^-_{\sigma,k}(t)]$,
one immediately finds that the dephasing rate is {\it temperature independent} and is given by 
\be
\label{gammas}
\Gamma_\sigma(t) = \frac{-1}{t}\sum_k \ln\left(1-2\frac{\eta^2_k}{\delta^2_k}\sin^2\delta_kt\right).
\ee

\noindent
(ii) {\it Bosonic bath:}
In the spin-boson model one obtains a closed-form expression for the time-evolution operator using the Magnus expansion, which truncates to second order as the commutator of the bosonic creations and annihilation operators is a $c-$number (App. A). 
Equations (\ref{uop})-(\ref{coh}) hold true even for a bosonic bath, if $U^\pm_\sigma$ are replaced with appropriate displacement operators for the oscillators,
\be
U^\pm_{b}(t) = \prod_k U^\pm_{b,k}(t),~~ U^\pm_{b,k}(t)={\rm e}^{i(\alpha_k(t)b_k-\alpha^*_k(t)b^\dagger_k)},
\ee
where $\alpha = (1-{\rm e}^{i\omega_kt})/\omega_k$. Here the eigenfrequencies of the bath particles are not renormalized (dressed) by the coupling unlike for the spin-bath.

Substituting the expression above in Eq. (\ref{coh}), we find that a bosonic bath would lead to temperature-dependent dephasing rate
\be
\label{gammaboson}
\Gamma_b(t)=\frac{1}{t}\sum_k \frac{\eta^2_k}{\omega^2_k}\coth\beta\omega_k\sin^2\omega_kt.
\ee

At low temperatures one always finds that
\be
\Gamma_\sigma(t) \gg \Gamma_b(t)
\ee
i.e., a spin bath leads to more harmful effects than a bosonic bath. In Fig. 1(a) we compared these rates for a Lorentzian bath.

It can be seen from Eq. (\ref{gammas}) that the spin-bath leads to complete dephasing at certain times if there exists even a single mode whose coupling strength satisfies $\eta_k \ge \omega_k$. This is quite dramatic as compared to the oscillator bath which cannot lead to such complete dephasing even if all the bath modes satisfy the condition $\eta_k \ge \omega_k$.

The other significant difference is the faster dynamics that the spin-bath induces due to its renormalized frequencies 
\be 
\delta_k \equiv \sqrt{\omega^2_k+\eta^2_k} > \omega_k.
\ee
Because of this faster dynamics, despite a similar bath coupling spectrum, a given  time scale in the Zeno regime for spin-boson dynamics can lie in the anti-Zeno regime for the spin-spin dynamics. This can make a huge difference to the control schemes which strongly rely on the appropriate time scales for operation.
In addition, the effect of renormalized frequencies becomes more evident in situations where the dominant coupling of the qubit is to the zero-frequency modes ($1/f$ noise \cite{fnoise}). In such cases the bosonic bath leads to slow exponential decay even at short times (Markovian dynamics), while the spin-bath can still lead to faster Gaussian decay (non-Markovian dynamics).

\begin{figure*}[htb]
\includegraphics[width=7cm]{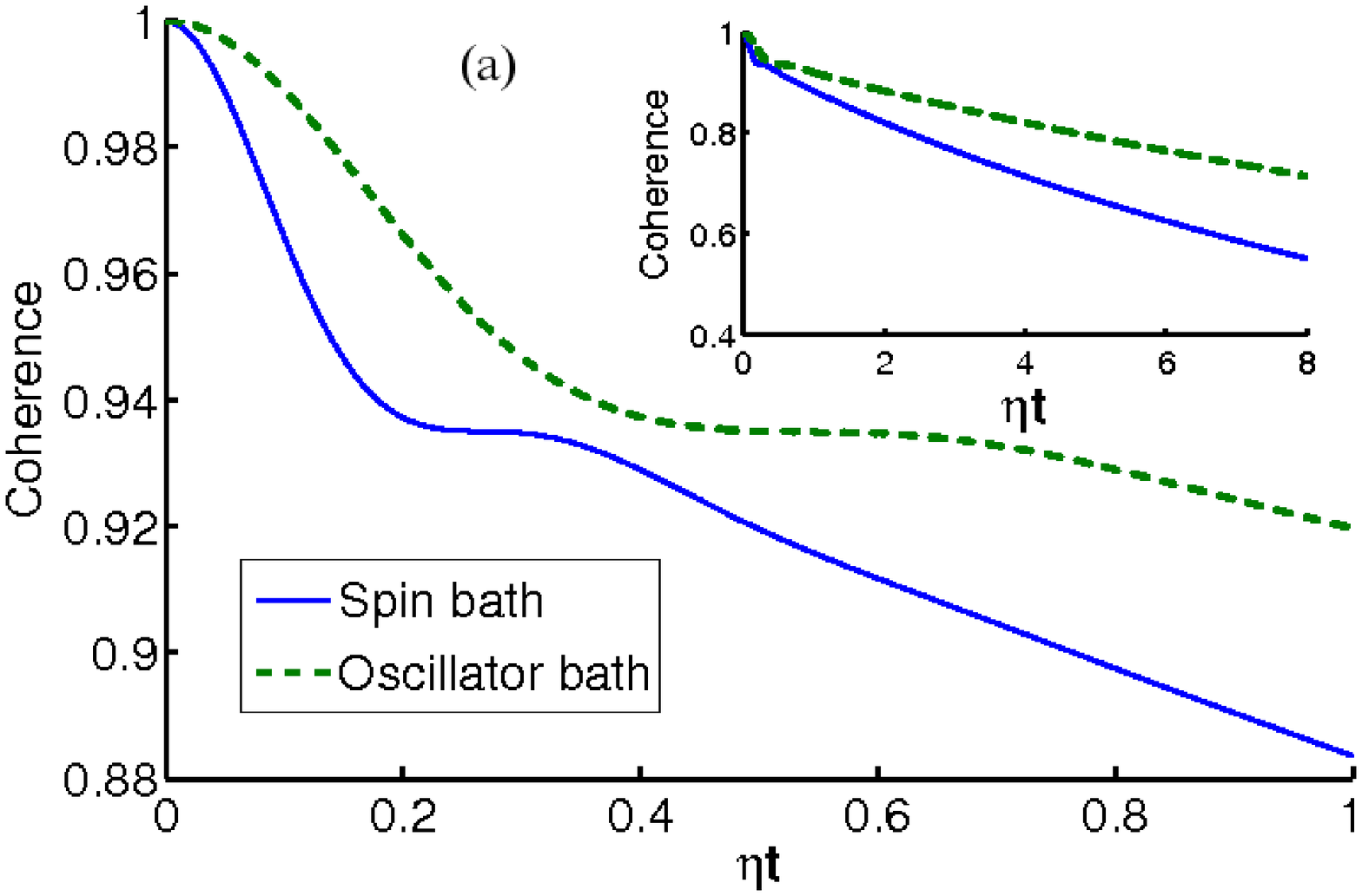}
\includegraphics[width=7cm]{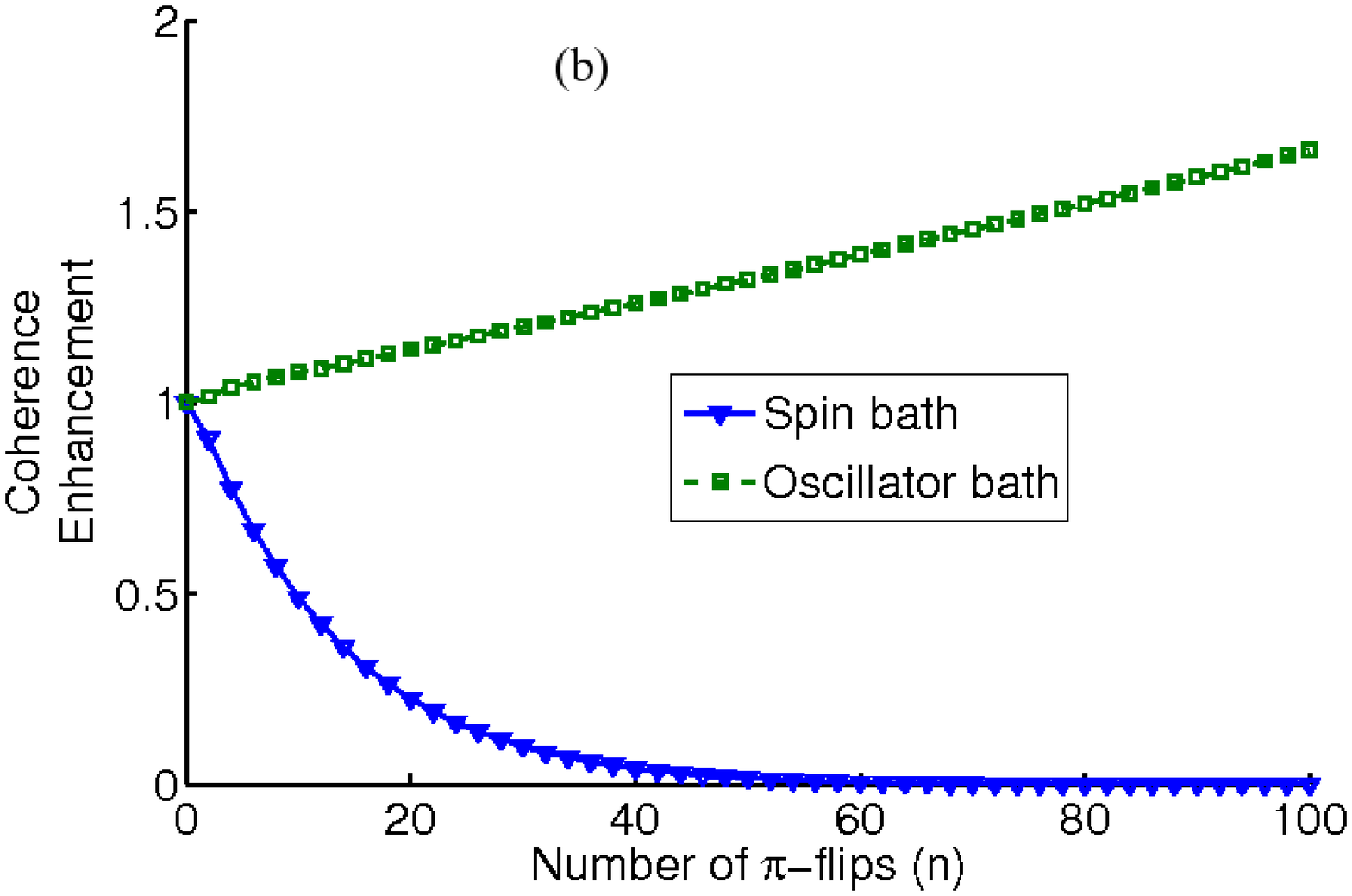}
\protect\caption
{{\bf Coherence dependence on quantum-statistics for qubit coupled to a Lorentzian bath}.{\bf(a)}. The free evolution of the qubit coherence $\rho^{+-}_S(t)$ is shown to be faster when the bath consists of  are either spins (solid-blue line) than when it consists of oscillators (dashed-green line). Time is in coupling strength units $\eta^{-1}$. In the inset shown is the evolution for longer times. {\bf(b)}. We have compared the relative change (enhancement) of qubit coherence in the presence of control with respect to its free evolution for spin and oscillator baths, under periodic $\pi$-pulses applied at a rate $\tau = T/n$, where $T$ is the total time and $n$ is the total number of pulses. }
\end{figure*}
\subsection{Dephasing control for Spin and Boson baths}
We consider brief control pulses along $\hat{x}$ at intervals $\tau$ as a means of dynamically decoupling the qubit from the bath. The simplest control that can significantly reduce the dephasing rate can be realized by applying frequent brief $\pi$-pulses in a Bloch-sphere direction $(\hat{x})$ orthogonal to the bath-interaction ($\hat{z}$).

\noindent
(i) {\it Bosonic bath:}
A closed form equation for $\Gamma_B(t)$ can be obtained for periodic $\pi$-pulses, 
\be
\label{picontb}
\Gamma_b(t)=\frac{1}{t}\sum_k \frac{\eta^2_k}{\omega^2_k}\coth\beta\omega_k{F}^n_{b,k}\sin^2\omega_kt,~{F}^n_{b,k}(\tau) = \frac{\sin^2\left(\frac{\omega\tau}{2}\right)\sin^2\left(\frac{n(\pi+\omega\tau)}{2}\right)}{\sin^2\left(\frac{\pi+ \omega\tau}{2}\right)\sin^2\left(\frac{n\omega\tau}{2}\right)}.
\ee
For bosonic baths one can obtain exact analytical expressions for the dephasing rates even for aperiodic $\pi$-pulses. In particular with high-frequency cutoff, an aperiodic pulse interval $\tau_n = T\sin^2\left(\frac{n\pi}{N}\right)$, where $T$ is the total time and $N$ is the total number of pulses, was shown to be the optimal pulse sequence for decoherence control \cite{uhrig1}. Yet, this pulse sequence may not be optimal if there is a constraint on the pulse number (energy constraint) \cite{jens}. Since the minimal pulse interval is determined by the total number of pulses $N$, the choice of $N$ determines whether these pulses give rise to  the Zeno or anti-Zeno effects \cite{koff}.

\noindent
(ii) {\it Spin bath:}
The form of $\Gamma_\sigma(t)$ in Eq. (\ref{gammas}) makes it difficult to obtain a closed-form expression when brief $\pi$ pulses are applied at aperiodic intervals $\tau_n$. Yet, the nature of spin-$1/2$ nature of the bath constituents allows one to exactly diagonalize the system-bath dynamics numerically for any number of bath particles in the presence of control. The control dynamics is determined by multiplying many $2\times 2$ matrices, which is evident from the structure of the unitary evolution operator:
\be
\label{sbcnt}
U_\sigma(t) = \left[\cdots U^+_\sigma(\tau_3)U^-_\sigma(\tau_2)U^+_\sigma(\tau_1)\right]|\mathcal{+}\rangle\langle +|+\left[\cdots U^-_\sigma(\tau_3)U^+_B(\tau_2)U^-_\sigma(\tau_1)\right]|{\bf -}\rangle\langle -|
\ee

A simple closed-form expression for $\Gamma_\sigma(t)$ when $\pi$-pulses are applied periodically at intervals $\tau$, (see App. B) is found to be
\be
\label{piconts}
\Gamma_\sigma(t) = \frac{-1}{t}\sum_k \ln\left(1-2{F}^n_{\sigma,k}(\tau)\frac{\eta^2_k}{\delta^2_k}\sin^2\delta_kt\right), ~{F}^n_{\sigma,k}(\tau)=\left|\frac{\sin n\phi_{k}}{2\sin \phi_{k}}\right|.
\ee
where
\be
\phi_k = \cos^{-1}(1-\frac{2\omega^2_k}{\delta^2_k}\sin^2\delta_k t).
\ee
As discussed earlier, the faster dynamics induced by the spin-bath modifies the crossover from the Zeno to anti-Zeno regime for a given bath spectrum. We have shown this explicitly in Fig. 1(b) and Fig. 2 for the case of a qubit coupled to a Lorentzian bath spectrum, where the periodic $\pi$-pulse control reduces the dephasing rate for a bosonic bath, while the same control enhances the dephasing rate for a spin bath. On the other hand, as the control pulse interval becomes shorter, the qubit dephasing is more strongly suppressed in the case of a spin-bath than in the case of an oscillator bath (see Fig. 2). This contrasting behavior stems from the faster dynamics in the presence of a spin-bath, due to dressing of the bath frequencies by the coupling strengths. 

\begin{figure*}[htb]
\includegraphics[width=12cm]{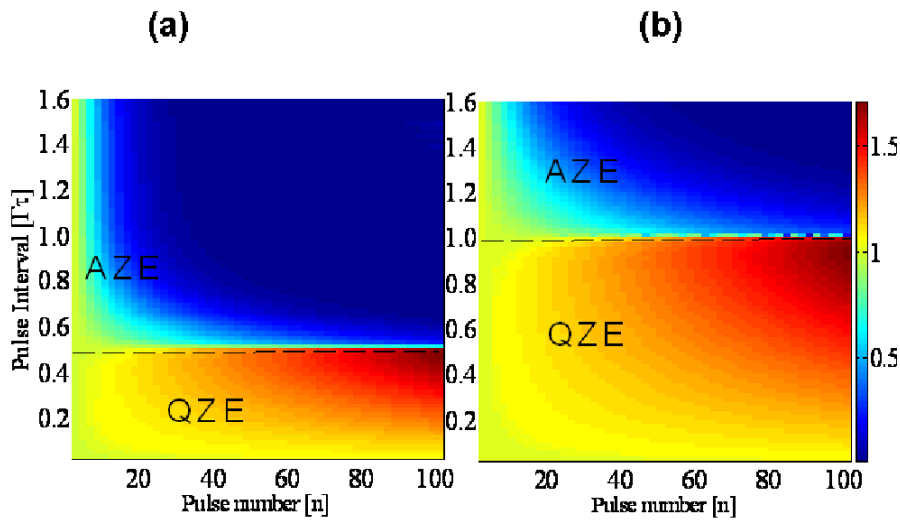}
\protect\caption
{{\bf Quantum Zeno (QZE) to anti-Zeno (AZE) cross over for qubit coupled to a Lorentzian bath under periodic $\pi$-pulses}. The relative change (enhancement) of qubit coherence in the presence of control with respect to its free evolution for spin (left) and oscillator (right) baths, is plotted as a function of the pulse interval $\tau$ (in units of inverse bath correlation time, $\Gamma$) and pulse number $n$. The transition from QZE (coherence enhancement) to AZE (coherence suppression) for bosonic baths is exactly at the bath correlation time $1/\Gamma$, while for spin baths it is well within the bath correlation time.}
\end{figure*}
\section{Weak coupling limit}
For boson baths the second-order (time-local) non-Markovian master equations \cite{koff} are known to be adequate for evaluating the qubit's decoherence rates and its modifications in the presence of control. This is true even in the limit of strong system-bath couplings, due to the commutator of the bosonic creation and annihilation operators being a $c$-number (App-I). As the spin-bath operators do not allow for this simplification it is instructive to discuss the limitations of the second-order analysis in the system-bath coupling strength.

\noindent
(i) {\it Spin bath:}
The  differential master equation (ME) for the density operator of the system $\rho_S(t)$, to second order in the system-bath coupling, is given by \cite{koff}
\be
\dot{\rho}_S(t) = -i[H_S(t),\rho_S(t)]+\int^t_0 dt^\prime \lbrace\Phi(t-t^\prime)[S(t)\rho_S(t),S(t^\prime)]+h.c.\rbrace.
\ee
The bath response function $\Phi(t-t^\prime) =\sum_k\eta^2_k \left[\cos(\omega_k(t-t^\prime))+i\langle\sigma_k\rangle\sin(\omega_k(t-t^\prime))\right]$ and $S(t) =U_S(t)S^zU^\dagger_S(t)$, where $U_S(t) = \exp(\int^t_0\Delta(t^\prime)dt^\prime S^x)$.
In the absence of control ($\Delta(t)=0$), the second order analysis leads to pure dephasing, at a rate
\be
\Gamma^{ME}_\sigma(t) = \sum_k \frac{2\eta^2}{\omega^2_k}(\omega_kt - \sin\omega_k t).
\ee
On comparing this expression with the exact form for a spin bath, Eq. (\ref{gammas}), one can see that the only similarity is the temperature-independent dephasing and $O(t^2)$ dynamics. Clearly, the second-order analysis overestimates the decay rate and does not capture the effect of renormalized system-bath couplings, as obtained from the exact analysis. 

Under periodic $\pi$-flip control, Eq. (\ref{piconts}) is replaced by
\be
\Gamma^{ME}_\sigma(t) = \sum_k \frac{2\eta^2}{\omega^2_k}{F}^n_{b,k}(\omega_kt - \sin\omega_k t),
\ee
where the expression for ${F}^n_{b,k}$ is given in Eq. (\ref{picontb}). 

The validity condition of second-order analysis for spin baths is stringent, i.e., 
\be \eta_k/\omega_k \ll 1
\ee

\noindent
(ii) {\it Bosonic bath}:
For a bosonic bath, the second-order analysis leads to an identical dephasing rate for both strong and weak coupling, as that obtained from the exact analysis, viz.,
\be
\Gamma^{ME}_b(t) = \frac{1}{t}\sum_k \frac{\eta^2_k}{\omega^2_k}\coth\beta\omega_k\sin^2\omega_kt.
\ee
Similarly, under periodic pulse pulse control, the expression for $\Gamma_B(t)$ is identical to that obtained under exact analysis (see Eq. (\ref{picontb})).

For the second-order analysis to be valid, coupling to a bosonic bath only needs to satisfy 
\be 
\eta_k/\omega_k \le 1.
\ee

\section{Conclusions}
In conclusion we have shown that the temperature-independence of the dephasing rate and the ``dressing'' of the bath frequencies by their coupling strengths render the dependence of the dephasing of a qubit by a spin bath very different from that caused by the well studied bosonic (oscillator) bath. Consequently, the quantum statistics of the bath determine the crossover from the Zeno to the anti-Zeno regime. This in turn affects the decoherence control: pulse timings required for Zeno-like dynamical decoupling depend on whether the bath is composed of spins or oscillators. Since pulse errors commonly increase monotonically with the number of pulses, there is a lower bound on $\tau$ in any experimental scenario. The useful pulse intervals for coherence control will therefore crucially depend on whether the qubit is coupled to a spin or bosonic bath. In particular, short-interval (high-rate) pulses can cause the leakage of the population out of the qubit subspace to higher levels of the system \cite{jens}.

\begin{acknowledgments}
We acknowledge the support of the EC (MIDAS), DIP, GIF, and ISF. 
 \end{acknowledgments}

\begin{appendix}
\section{Magnus Expansion}
To obtain a closed-form expression for the time-ordered unitary operator 
\be
U(t)=T_\leftarrow \exp\left[-i\int_0^t dt'H_I(t')\right],
\ee
we resort to the Magnus expansion \cite{magnus} of the exponent of $U(t) = \exp(\Omega(t))$. 
The first few terms of the expansion are
\bea
\Omega(t) &=& \int^t_0 H_I(t_1)dt_1 - \frac{1}{2}\int^t_0\int^{t_1}_0[H_I(t_1), H_I(t_2)]dt_1dt_2 \nn \\ &&+\frac{1}{6}\int^t_0\int^{t_1}_0\int^{t_2}_0dt_1dt_2dt_3([H_I(t_1),[H_I(t_2),H_I(t_3)] + [[H_I(t_1),H_I(t_2)],H_I(t_3)])+\cdots.
\eea

\noindent
(i) {\it Bosonic baths:}
Let us consider Eq. (1) of the main text, and set $\Delta(t)=0$.
In the interaction picture this yields
\be
H_I(t) = {\rm e}^{iH_Bt}H_I{\rm e}^{-iH_Bt} = S_z\sum_k\eta_k({\rm e}^{-i\omega_kt}b_k+{\rm e}^{-i\omega_kt}b^\dagger_k).
\ee
For bosonic bath operators, the commutator of the interaction Hamiltonian at two different times is a $C$-number function in the bath operators:
\be
[H_I(t),H_I(t')]=-2i\sum_k \eta^2_k \sin \omega_k (t-t'). \nn
\ee
The fact that this commutator is a $C$-number implies that only the first two terms of the expansion are non-zero. Now, the closed-form equation for the time-evolution operator takes the simple form
\bea
\label{uexp}
U(t) &=& \sum_{\pm} U_\pm(t)|\pm\rangle\langle\pm| \nn \\
U_\pm(t) &=& \exp\left[{-itf(t)} \pm \sum_k\left(\alpha_k(t)b^\dagger_k-\alpha^*_k(t)b_k\right)\right].
\eea
The coupling to the bath determines the functions
\be
f(t)=\frac{1}{t}\sum_k \eta^2_k(\omega_k t-\sin \omega_kt)/\omega^2_k, ~~\alpha_k(t) =  \eta_k\frac{1-{\rm e}^{i\omega_kt}}{\omega_k}.
\ee

\noindent
(ii) {\it Spin baths:} Similar to the above analysis, in the interaction picture we get
\be
H_I(t) = {\rm e}^{iH_Bt}H_I{\rm e}^{-iH_Bt} = S_z\sum_k\eta_k({\rm e}^{-i\omega_kt}\sigma^-_k+{\rm e}^{-i\omega_kt}\sigma^+_k).
\ee
For spin bath operators, the commutator of the interaction Hamiltonian at two different times is an operator:
\be
[H_I(t),H_I(t')]=-2i\sum_k \sigma^z_k \eta^2_k \sin \omega_k (t-t'). \nn
\ee
The fact that this commutator is a not a $C$-number implies that all terms of the expansion are non-zero and it is quite complex to obtain a closed-form equation for the time-evolution operator using magnus expansion.

\section{Periodic $\pi$-pulse control for spin baths}
For $2n$ periodic $\pi$-pulses, Eq. (\ref{sbcnt}) gets simplified to 
\be
\label{app1}
U(t) = \left[U^-_B(\tau)U^+_B(\tau)\right]^n|\mathcal{+}\rangle\langle +|+\left[U^+_B(\tau)U^-_B(\tau)\right]^n|{\bf -}\rangle\langle -|.
\ee
Hence a closed form expression for $U(t)$ can be obtained by diagonalizing the matrix $\left[U^-_B(\tau)U^+_B(\tau)\right]^n$.
Since
\be
\label{app2}
\left[U^-_B(\tau)U^+_B(\tau)\right]^n = \prod_n \left[U^-_{k,B}U^+_{k,B}\right]^n,
\ee it is sufficient to diagonalize the effect of one bath spin on the qubit after $2n$ pulses.

Since the eigenvectors of a unitary matrix $U^-_{k,B}U^+_{k,B}$ are orthogonal, evaluating arbitrary powers of this matrix becomes quite simple.
\bea
U^-_{k,B}U^+_{k,B} &=& \left(\cos^2\delta_kt -\frac{\omega^2_k-\eta^2_k}{\delta^2_k}\sin^2\delta_kt\right)\hat{\rm I}+\frac{2i\omega_k\sin\delta_kt}{\delta^2_k}\left(\delta_k\cos\delta_kt\sigma^z_k-\eta_k\sin\delta_kt\sigma^y_k\right) \nn \\
&=&\lambda^+_k|v^+_k\rangle\langle v^+_k| + \lambda^-_k|v^-_k\rangle\langle v^-_k|.
\eea
The corresponding eigenvalues and eigenvectors are given by
\bea
\lambda^\pm_k &=& x_k\pm i\sqrt{1-x^2_k}, \nn \\
|v^\pm_k\rangle &=&\frac{1}{\sqrt{1+\alpha^2_k}}[|0\rangle_k \mp i\alpha_k|1\rangle_k], ~\alpha = \frac{\sqrt{1-x^2_k-y^2_k}}{\sqrt{1-x^2_k}+y_k},
\eea
where 
\be 
x_k = 1-\frac{2\omega^2_k}{\delta^2_k}\sin^2\delta_kt, ~y_k = \frac{\omega_k}{\eta_k}\sin2\delta_kt.
\ee

Now using the fact that $\langle v^+_k|v^-_k \rangle = 0$ and $\langle v^\pm_k|v^\pm_k \rangle = 1$,
\be
\left[U^-_{k,B}U^+_{k,B}\right]^n = (\lambda^+_k)^n|v^+_k\rangle\langle v^+_k| + (\lambda^-_k)^n|v^-_k\rangle\langle v^-_k|.
\ee

Substituting this in Eq's (\ref{app2}), (\ref{app1}) and performing the trace over bath degrees of freedom, one can obtain the dynamically modified dephasing rate of the qubit,
to be
\be
\label{app3}
\Gamma_S(t) = \frac{-1}{t}\sum_k \ln\left(1-2{F}^n_{\sigma,k}(\tau)\frac{\eta^2_k}{\delta^2_k}\sin^2\delta_kt\right), ~{F}^n_{\sigma,k}(\tau)=\left|\frac{\sin n\phi_{k}}{2\sin \phi_{k}}\right|.
\ee
where
\be
\phi_k = \cos^{-1}(1-\frac{2\omega^2_k}{\delta^2_k}\sin^2\delta_k t).
\ee
\end{appendix}

\end{document}